\documentclass[useAMS,usenatbib]{mn2e}

\def\apjl{ApJL}
\def\mnras{MN}
\def\apj{ApJ}
\def\aj{AJ}

\usepackage{graphicx}
\usepackage{float}
\usepackage{amssymb}
\usepackage{amsmath} 
\usepackage{mathrsfs} 

\title[The core size of the Fornax dwarf]{The core size of the Fornax dwarf Spheroidal} 
\author[N. C. Amorisco, A. Agnello and N. W. Evans]{N. C. Amorisco$^{1}$\thanks{E-mail:
    amorisco@ast.cam.ac.uk, aagnello@ast.cam.ac.uk, nwe@ast.cam.ac.uk}, A. Agnello$^{1}$ and N. W. Evans$^{1}$\\ $^{1}$Institute of Astronomy, University of
  Cambridge, Madingley Road, Cambridge CB3 0HA, UK}

\begin{document}

\date{Accepted, Received }

\pagerange{\pageref{firstpage}--\pageref{lastpage}} 

\maketitle

\label{firstpage}

\begin{abstract}
We exploit the detection of three distinct stellar subpopulations in
the red giant branch of the Fornax dwarf Spheroidal to probe its
density distribution. This allows us to resolve directly the evolution
with radius of the dark matter mass profile.  We find that a cored
dark matter halo provides a perfect fit to the data, {being consistent with 
all three stellar populations well within 1-sigma,} and for the first
time we are able to put constraints on the core size of such a halo. With
respect to previous work, we do not strengthen the statistical
exclusion of a dark matter cusp in Fornax, but we find that
Navarro-Frenk-White haloes would be required to have unrealistically
large scale radii in order to be compatible with the data, hence low
values of the concentration parameter. We are then forced to conclude
that the Fornax dwarf Spheroidal sits within a dark matter halo having
a constant density core, with a core size of $r_0=1^{+0.8}_{-0.4}$kpc.
\end{abstract} 

\begin{keywords}
galaxies: kinematics and dynamics -- Local Group -- galaxies:
individual; Fornax dSph
\end{keywords}
%

\section{Introduction}

Dynamical modelling of the dwarf Spheroidals (dSphs) has now been
tackled with a number of different techniques. At the heart of these
efforts lies the importance of constraining the properties of the
central dark matter density profile, so as to advance our
understanding of galaxy formation towards the low-mass end of the
mass-spectrum. Jeans equation analyses~\citep[e.g.][]{Wa09, Lo09},
phase-space analyses~\citep{Wi02, Wu07, Am11} as well Schwarzschild
modelling~\citep{Ja12, Bre12} have all now been used in this field,
with varied degrees of success.  Nonetheless, it is fair to say that
the main step forward has not really come through brute force in the
modelling techniques, but rather from the realization that the
presence of multiple stellar populations in dSphs could be turned into
a powerful strength.

After the first work of~\citet{Ba08}, the method has now been
amplified to overcome some of the deficiencies of Jeans
analyses~\citep{Ev09, Am12a} and to give the required robustness to
the identification and disentanglement of the coexisting stellar
subpopulations~\citep[][ WP11 in the following]{WaP11}.  Very recently,
\citet{Ag12} (AE12) have brought the argument into sharp focus.  They
show that the additional evidence coming from coexisting
subpopulations is extremely simple in nature, and can be fully
understood in terms of the independent requirements to the global
energetics of the system enforced by the different stellar
subpopulations. By using the projected virial theorem (PVT) it is
possible to identify the dark matter profiles that are compatible with
the photometry and kinematics of the subpopulations and exclude those
in which the subpopulations would not all be in equilibrium at the
same time.

New rich and precise datasets will certainly allow and justify full
use of more refined dynamical techniques in the near future,
especially exploiting the spatially resolved properties of the line of
sight (LOS) velocity distribution~\citep[][ JG12 in the
following]{Am12b, Ja12}.  On the other hand, while the detailed
orbital distribution of the stars is still a major uncertainty in the
modelling, it is helpful to work within a framework that is entirely
free from any anisotropy-driven degeneracy, such as the PVT.

All the analyses mentioned so far point towards the presence of a
constant density core at the center of both studied dSphs, Sculptor
and Fornax. Together with previous evidence pertaining to low surface
brightness galaxies~ \citep[e.g.][]{deB01}, these results have
triggered a number of both theoretical and numerical studies.  The aim
is of course to understand the complex baryonic processes responsible
for transforming the initial dark matter $\rho\sim r^{-1}$ cusp
predicted by Cold Dark Matter (CDM) models~\citep{Du91, Na97}
\begin{equation}
\rho_{\rm NFW}(r)={\rho_0\over{\left({r\over{r_0}}\right)\left(1+{r\over{r_0}}\right)^2}}
\label{nfw}
\end{equation}
into a core. A number of suggestions have been made, including
multiple epochs of mass loss and gas re-accretion~\citep{Re05}, clumpy
baryonic infall~\citep{Co11} and intermittent and concentrated star
formation and subsequent supernova feedback~\citep{Po12}. Some or all
of these processes may be able to deposit sufficient energy in the
dark matter halo to erase the central cusp. This picture has been
tested with promising results by~\citet{Tey12} and~\citet{Br12}, who
show that a more realistic treatment of star formation and feedback
can reduce the tension between CDM and the detection of cores in
luminous dSphs.

In this {\it Letter}, we revisit the dark matter density profile of the
Fornax dSph, after~\citet{Am12c} (A12) found that a three-population
division better describe the data with respect to a two-population
division (WP11). Both previous analyses of the Fornax dSph (WP11,
JG12) were able to exclude the presence of an Navarro-Frenk-White
(NFW) cusp with high statistical significance, although the size of
the constant density core has remained poorly constrained. WP11 obtain
a measurement of the average logarithmic slope of the mass profile
$\Gamma$ (see eqn.~(\ref{Gamma})) at a radius that is comparable with
the half-light radii of their two subpopulations. This only allows
them to put a lower limit to the core size. Similarly, JG12 find that
a constant density core is required, but orbit-based modelling {of a single stellar population} would
require a more extended set of kinematic tracers in order to put an
upper limit to its scale length. By exploiting all the three different
stellar populations in Fornax, we are able to {find} such a
constraint.

\section{Probing the mass profile}

A12 present evidence for the coexistence of three distinct stellar
subpopulations in the red giant branch of the Fornax dSphs. These
subpopulations have different metallicities, spatial distributions and
kinematics. We refer to their Table~1 for details, and record here for
convenience the relevant half-light radii:
\begin{equation}
\left\{\begin{array}{ccc}
R_{\rm h}^{\rm MP}&=&935\pm 65 {\rm pc},\\
R_{\rm h}^{\rm IM}&=&610\pm 25 {\rm pc},\\
R_{\rm h}^{\rm MR}&=&437\pm 55 {\rm pc},
\end{array}\right.
\label{hlr}
\end{equation}
where MP, IM and MR stand, respectively, for metal-poor,
intermediate-metallicity and metal-rich. Fig.~1 displays the LOS
velocity dispersion profiles of the three stellar
sub-populations. These have been obtained by using the maximum
likelihood technique described in~\citet{Am12b}, which takes into
account the probabilities of membership of each star to the identified
stellar subpopulations.
\begin{figure}
\centering
\includegraphics[width=\columnwidth]{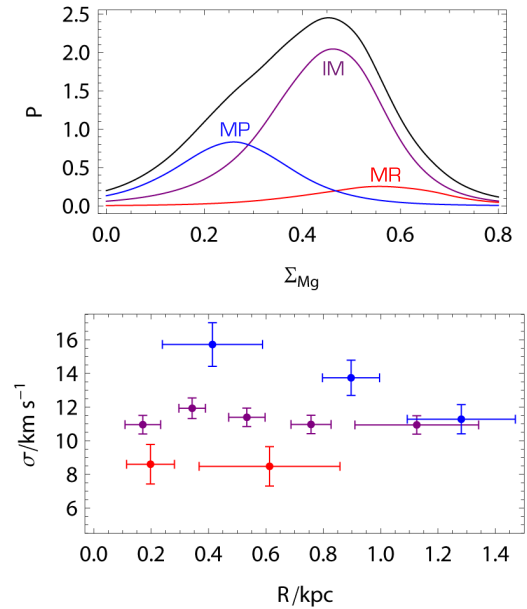}
\caption{Metallicity distribution (in the Magnesium index -- in \AA) and line of
  sight velocity dispersion profiles in circular annuli for the three
  stellar subpopulations identified in Fornax. Color-coding
  corresponds to increasing metallicity. {The profiles use the same
  number of tracers per bin, although this is different for each
  population (34 for MR, 283 for IM and 123 for MP).} }
\label{kinprofs}
\end{figure}
\begin{figure*}
\centering
\includegraphics[width=.85\textwidth]{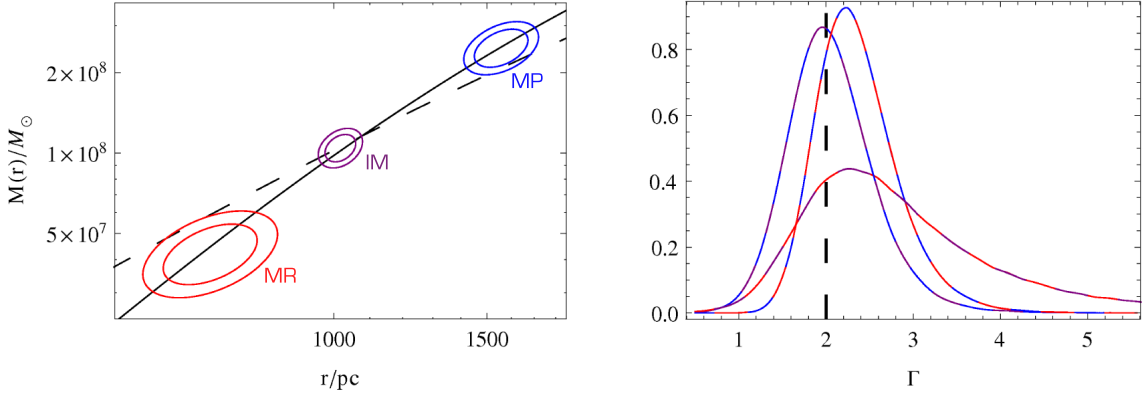}
\caption{Left panel: estimates of the total enclosed mass $M(r)$ as
  obtained from each of the three stellar subpopulations; color-coding
  is the same as in Fig.~1. Right panel: probability distributions of
  the logarithmic slope $\Gamma$ obtained by pairing the mass
  estimates in the left panel (colors are paired accordingly).}
\label{blobs}
\end{figure*}

With this information,we produce a total mass estimate for each
subpopulation.  We use the mass estimator proposed by~\citet{Am12a}
\begin{equation}
M\left[(1.67\pm0.04)R_{\rm h}\right]=(5.85\pm0.2)\ {{R_{\rm h}\ \sigma^2_{\rm los}(R_{\rm h})}\over G}\ ,
\label{massest}
\end{equation}
which has been proved to be accurate on the wide family of models with
Michie-King phace space distribution. We also notice that use of
alternative mass estimators with the same functional structure -- as
for example the one proposed by~\citet{Wa09} and used in~WP11 -- would
give identical results for the logarithmic slope of the mass profile
\begin{equation}
\Gamma_{i,j}={{\log M_i-\log M_j}\over{\log R_{\rm h, i}-\log R_{\rm h, j}}}\ ,
\label{Gamma}
\end{equation}
where $i$ and $j$ identify any two stellar subpopulation. The left
panel of Fig.~2 illustrates the one and two sigma contours for each of
the total mass estimates, with the same color-coding as in Fig.~1; the
right panel of the same Figure shows the probability distributions of
the three logarithmic slopes obtained by combining in pairs the
distinct subpopulations.

For any realistic dark matter density profile, the logarithmic slope
$\Gamma(r)$ is a monotonic function of radius and decreases to zero at
large radii.  For a cored halo, $\Gamma\lesssim 3$ within the scale
radius of the core.  For a cusped dark matter profile $\rho\sim
r^{-\gamma}$, $\Gamma\lesssim 3-\gamma$ near the center. Therefore,
for an NFW halo, $\Gamma$ is never higher than 2.  WP11 have measured
a value of $\Gamma$ that is systematically higher than 2 near the
center of Fornax, and were able to exclude the presence of a cusp
($\gamma\geq 1$) with high significance ($s\geq96\%$).  Since we have
three different mass estimates, we can measure the logaritmic slope of
the mass profile in multiple locations, hence assess directly its
evolution with radius. In the radial interval defined by the IM and MP
stellar populations, which is the farthest from the center, the slope
is centered on $\Gamma\approx 2$ (blue-purple curve in the right panel
of Fig.~2).  However, as soon as the radius is reduced by considering
the MR stellar population, the mean value of $\Gamma$ sistematically
increases. By considering the MR-MP pair (red-blue curve) we obtain
$\langle\Gamma\rangle\approx 2.4$, while, near the center, the MR-IM
pair (red-purple curve) suggests $\langle\Gamma\rangle\approx 2.65$.
This is probing directly the radial variation of the mass profile of
the Fornax dwarf.

In a purely statistical sense, the exclusion of an NFW cusp we obtain
from this analysis is not as strong as the one obtained in WP11: the
tails of the probability distributions of both $\Gamma_{\rm MR, MP}$
and $\Gamma_{\rm MR, IM}$ suggest that the probability of a density
profile as steep or steeper than an NFW is $p\lesssim
22\%$. Nonetheless, it is worth noticing that such a cusped density
profile would be unrealistic for different reasons. For instance, the
dashed curve in the first panel of Fig.~2 represents the best fitting
NFW halo with scale radius $r_0=3$~kpc. The agreement with the mass
estimates is marginal and it is in fact impossible to obtain a better
agreement for any NFW density profile with a smaller scale radius
(hence even smaller $\langle\Gamma\rangle$ at the same radii).
Different studies of Fornax~\citep[][WP11]{Ir95, Co12} suggest that
its tidal radius is likely to be smaller than 3~kpc, hence presenting
an evident incompatibility with the required scale
radius. Furthermore, even an NFW halo with a scale radius of `only'
$r_0>2$~kpc would already have a concentration $c<9$, well below the
expectation of CDM models~\citep[e.g.][]{Ma07}.  As a comparison, the full
line in the left panel of Fig.~1 represents a Burkert
profile~\citep{Bu95}
\begin{equation}
\rho_{\rm Bur}(r)={\rho_0\over{\left(1+{r\over{r_0}}\right)\left[1+\left({r\over{r_0}}\right)^2\right]}}
\label{bur}
\end{equation}
with a core-size of $r_0\approx 1.4$kpc, which provides excellent
agreement with the mass estimates, {being consistent with 
all three stellar populations well within 1-sigma.}

\section{Projected Virial Theorem}

We can gain a more systematic insight by making use of the PVT
\begin{equation}
2K_{\rm los}+W_{\rm los}=0\ ,
\label{PVT}
\end{equation}
where, in the usual notation, $K_{\rm los}$ and $W_{\rm los}$ are the
projected components of the pressure and potential energy tensors --
explicit formule are recorded in AE12. The PVT provides a fundamental
means of exploring the energetics of a system with multiple stellar
subpopulations.  As in AE12, we assume that the photometry of the
different subpopulations is well represented by a Plummer profile; {we 
have performed the same analysis using exponential profiles and find consistent results.}

In the ideal case of no observational uncertainties, use of the PVT
proceeds as follows. For an assigned dark matter density profile, once
the surface brightness of a stellar subpopulation $\mu(r)$ and its
velocity dispersion profile $\sigma_{\rm los}(r)$ are given, the PVT
yields the characteristic density $\rho_0$ that needs to be coupled
with the scale radius $r_0$ in order to have an equilibrium
configuration. {In our application, we use a Monte Carlo procedure
to propagate any observational uncertainties in the half-light radius
of the photometry from eqn.~(\ref{hlr}) as well as in the kinematics
(see Fig.~1), and construct a complete probability distribution for
the characteristic density $\rho_0$ at any fixed characteristic radius
$r_0$.}

The left panel of Fig.~3 shows the one-sigma virial stripes that we
obtain when each subpopulation in embedded in an NFW dark matter
profile (eqn.~(\ref{nfw})); the right panel illustrates the case of a
Burkert halo (eqn~(\ref{bur})). In the NFW case, the one-sigma regions
related to the IM and MP subpopulations overlap before
$r_0=1$ kpc. This is in agreement with the previous Section, since we
found $\langle\Gamma_{\rm MR, IM}\rangle= 2.0$ at large radii.
However, at smaller radii the MR subpopulation does not fit into this
picture.  Any NFW halo that is compatible at the one-sigma level with
all three subpopulations has $r_0\gtrsim 2$kpc, with the difficulties
mentioned earlier. {On the other hand, for a constant density core, 
the three one-sigma stripes show a consistent overlap region.}

Use of a single subpopulation does not allow any inference on the
scale radius of the core if taken by itself. It is by multiplying the
probability distributions defined by the three virial stripes that we
can constrain the core size. Fig.~4 displays the one- and two-sigma
confidence regions associated with the joint likelihood, together with
the marginalized probability distributions for the characteristic
density $\rho_0$ and core size $r_0$. All full lines are associated
with a Burkert dark matter profile (eqn.~(\ref{bur})), but we find
that the data do not show any significant preference for other
functional prescriptions of the core.  For example, we have explored
the density profile {used} by AE12
\begin{equation}
\rho_{\rm cNFW}(r)={\rho_0\over{\left[1+\left({r\over{r_0}}\right)^2\right]^{3/2}}}
\label{cnfw}
\end{equation}
and we find no difference in either the virial stripes or in the
maximum value of the joint likelihood.  In fact, the dotted line in
the lower panel of Fig.~4 has been obtained by using the density
profile~(\ref{cnfw}), but this is virtually indistinguishable from the
one obtained for a Burkert profile. As a comparison, the dashed line
in the same panel is associated with the joint likelihood of an NFW
halo.  Increasing the scale radius $r_0$ is the only way to increase
the average value of $\Gamma$ in the radial region sampled by the
three stellar subpopulations, therefore the marginalized probability
distribution is a monotonically increasing (respectively, decreasing)
function of $r_0$ (of $c$). We then conclude that the 68\% confidence
interval for the core size is $r_0=1^{+0.8}_{-0.4}$kpc.

\begin{figure*}
\centering
\includegraphics[width=.95\textwidth]{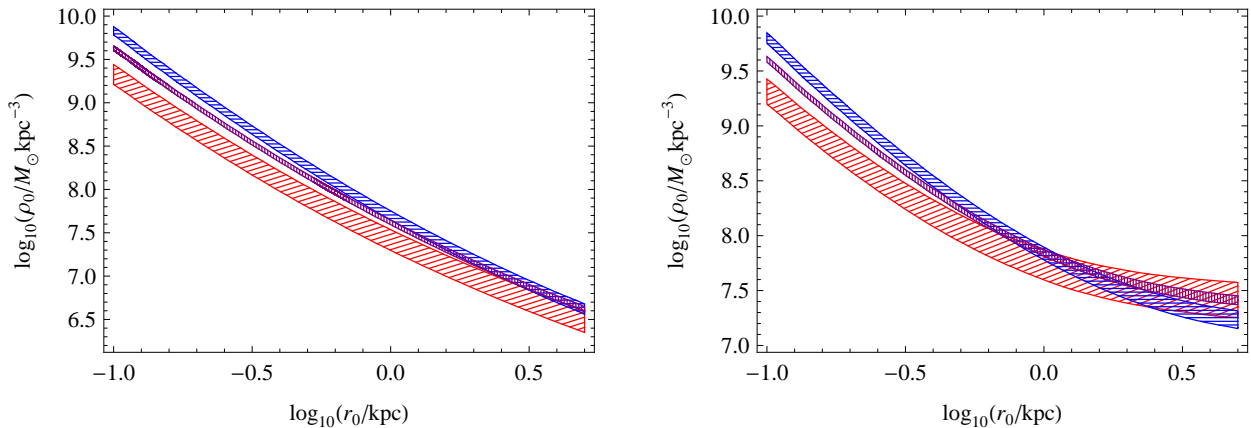}
\caption{Left panel: virial stripes for the three subpopulations
  embedded in an NFW gravitational potential. Right panel: virial
  stripes for the three subpopulations embedded in a Burkert
  gravitational potential. In both panels stripes are associated with
  1-$\sigma$ regions and color coding is the same as in Fig.~1 and~2.}
\label{kinprofs}
\end{figure*}
\begin{figure}
\centering
\includegraphics[width=.9\columnwidth]{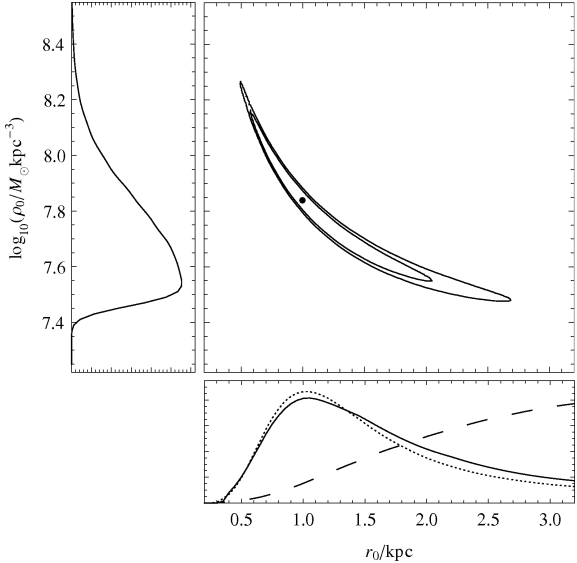}
\caption{Joint likelihood analisys of the virial stripes: one and two
  sigma contours in the halo plain $(r_0 ; \rho_0)$ and marginalized
  probability distributions for a Burkert density profile (full lines). {The short-dashed
  probability distribution is associated with the cored profile of eqn.~(7), while the
  long-dashed distribution is indicative of an NFW halo.}}
\label{jlik}
\end{figure}

\section{Conclusions}

The Fornax dwarf is the brightest, and probably the most massive of
all the dwarf spheroidal (dSph) companions surrounding the Milky Way
Galaxy. It has a complex evolutionary pathway, as is already evident
from its star formation history~\citep{Co08, deB12}. The detection of
localized overdensities~\citep{Co04, Co05} as well as a complex
dependence on metallicity of its rotational properties~\citep{Am12c}
support the idea of a relatively recent interaction with a smaller dwarf.
Large spectroscopic samples of thousands of
red giant stars in Fornax are now available~\citep{Wa09a}, and have
recently been separated into three populations -- metal-rich,
intermediate-metallicity and metal-poor -- by \citet{Am12c} using
chemo-dynamical methods. 

In stellar dynamics, multiple populations provide a powerful method to
constrain the common gravitational potential in which the stars move.
Here, we have exploited the division of Fornax into three populations
to show that the stars reside in a dark matter halo having a constant
density core with size of $r_0=1^{+0.8}_{-0.4}$kpc. This is in
contrast to the case when just the single, combined stellar population
of Fornax is studied. Then, the kinematic data are equally consistent
with cusps and cores~\citep{Ev09,St10}. Although often cited as an
issue, the ellipticity of the luminous dSphs is unable to alter the
conclusions on their dark matter density profiles, as shown by
\citet{Ag12}. {Even an oblate NFW halo would still need to be so
large and diffuse to be substantially different from expectations.}

All this adds weight to the evidence that the dSph galaxies of the Milky
Way possess dark matter cores. This evidence includes the survival of
kinematically cold substructure in Ursa Minor~\citep{Kl03}, the
longevity of the globular cluster population in Fornax~\citep{Go06, SS06, Co12},
as well as the kinematics of stars in Sculptor~\citep{Ba08,Am12a,Ag12}
and in Fornax~\citep{WaP11}. There is no direct evidence in favour of
a dark matter cusp in any dSph.

The nature of the central parts of dark haloes of dSphs is important
for two reasons. First, indirect detection experiments are monitoring
the sky for signals of the expected $\gamma$ ray emission produced by
pair annihilation of weakly interacting dark matter particles. For
example, the Large Area Telescope (LAT) on the Fermi satellite has
constrained the integral $\gamma$ ray flux from 14 dSphs to place
limits on the annihilation cross-section~\citep{Ab10, Ack11}. The beam-size
of the LAT is comparable to the angular half-light radius of most
dSphs ($\sim 0.5^\circ$). In turn, this is comparable to the angular scale
where uncertainties from the dark matter profile on the J factor 
are minimized~\citep[][]{Wa11, Ch11}. Therefore, whether cored or 
cusped models are used makes only modest differences
to the interpretation of the Fermi LAT data. However, arrays of
atmospheric Cerenkov telescopes like MAGIC and HESS
~\citep[e.g.,][]{Al11,Ab11} provide superior angular resolution, and
now the existence of a dark matter core rather than a cusp
substantially reduces the expected $\gamma$ ray
flux~\citep[e.g.,][]{Ev04}.

Secondly, it is a unique prediction of cold dark matter theories that
the centres of dark haloes are cusped. The dSphs of the Milky Way are
overwhelmingly dark matter dominated, and so provide natural testing
grounds for assessing this hypothesis.  The finding that the haloes of
at least some of the dSphs are cored may be interpreted in one of two
ways. Either it provides evidence for warm dark matter particles which
-- judging from the phase space density in the Fornax core -- have
mass $\sim 0.5$ keV~\citep{Da01}. Or, it provides evidence that
baryonic processes such as supernovae feedback or episodic mass loss
or infall have modified the pristine dark matter cusp. There is no
shortage of suggestions of processes that could eradicate the cusps
\citep[see e.g.][]{Re05,Co11,Po12}. The increasing size and richness
of datasets on the dSphs, combined with the increasing sophistication
of analysis tools, gives hope to the suggestion that observational
data may soon allow us to discriminate between some of these
possibilities.

\section*{Acknowledgments}
NA and AA thank the Science and Technology Facility Council and the
Isaac Newton Trust for financial support. We thank the anonymous
referee for a critical reading of the manuscript.


\begin{thebibliography}{99}

\bibitem[Abdo et al.(2010)]{Ab10} Abdo, A.~A., Ackermann, 
M., Ajello, M., et al.\ 2010, \apj, 712, 147 

\bibitem[Abramowski et al.(2011)]{Ab11}Abramowski, A., Acero, F., et al.\ 2011, 
Astroparticle Physics, 34, 608 

\bibitem[Agnello 
\& Evans(2012)]{Ag12} Agnello, A., \& Evans, N.~W.\ 2012, \apjl, 754, L39

\bibitem[Ackermann et al.(2011)]{Ack11} Ackermann, M., 
Ajello, M., Albert, A., et al.\ 2011, Physical Review Letters, 107, 241302 

\bibitem[Aleksi{\'c} et al.(2011)]{Al11} Aleksi{\'c}, J., Alvarez,
  E.~A., Antonelli, L.~A., et al.\ 2011, JCAP, 6, 35

\bibitem[Amorisco 
\& Evans(2011)]{Am11} Amorisco, N.~C., \& Evans, N.~W.\ 2011, \mnras, 411, 2118 

\bibitem[Amorisco 
\& Evans(2012a)]{Am12a} Amorisco, N.~C., \& Evans, N.~W.\ 2012a, \mnras, 419, 184 

\bibitem[Amorisco 
\& Evans(2012b)]{Am12b} Amorisco, N.~C., \& Evans, N.~W.\ 2012b,
  \mnras, in press (arXiv:1204.5181)

\bibitem[Amorisco 
\& Evans(2012c)]{Am12c} Amorisco, N.~C., \& Evans, N.~W.\ 2012c, ApJL,
  in press (arXiv:1206.6691)

\bibitem[Battaglia et al.(2008)]{Ba08} Battaglia, G., Helmi, 
A., Tolstoy, E., et al.\ 2008, \apjl, 681, L13 

\bibitem[Breddels et al.(2012)]{Bre12} Breddels, M. et al.\ 2012, arXiv:1205.4712

\bibitem[Brooks 
\& Zolotov(2012)]{Br12} Brooks, A.~M., \& Zolotov, A.\ 2012, arXiv:1207.2468 

\bibitem[Burkert(1995)]{Bu95} Burkert, A.\ 1995, \apjl, 447, 
L25 

\bibitem[Charbonnier et al.(2011)]{Ch11} Charbonnier, A., 
Combet, C., Daniel, M., et al.\ 2011, \mnras, 418, 1526 

\bibitem[Cole et al.(2011)]{Co11} Cole, D.~R., Dehnen, W., \&
  Wilkinson, M.~I.\ 2011, \mnras, 416, 1118

\bibitem[Cole et al.(2012)]{Co12} Cole, D.~R., Dehnen, W., Read,
  J.~I., \& Wilkinson, M.~I.\ 2012, arXiv:1205.6327

\bibitem[Coleman et al.(2004)]{Co04} Coleman, M., Da Costa, G.~S.,
  Bland-Hawthorn, J., et al.\ 2004, \aj, 127, 832

\bibitem[Coleman et al.(2005)]{Co05} Coleman, M.~G., Da 
Costa, G.~S., Bland-Hawthorn, J., \& Freeman, K.~C.\ 2005, \aj, 129, 1443 

\bibitem[Coleman \& de Jong(2008)]{Co08} Coleman, M.~G., \& de Jong,
  J.~T.~A.\ 2008, \apj, 685, 933

\bibitem[de Blok et al.(2001)]{deB01} de Blok, W.~J.~G., 
McGaugh, S.~S., Bosma, A., \& Rubin, V.~C.\ 2001, \apjl, 552, L23 

\bibitem[Dalcanton \& Hogan(2001)]{Da01} Dalcanton, J.~J., \& Hogan,
  C.~J.\ 2001, \apj, 561, 35

\bibitem[de Boer et al.(2012)]{deB12} de Boer, T.~J.~L., 
Tolstoy, E., Hill, V., et al.\ 2012, arXiv:1206.6968 

\bibitem[Dubinski \& Carlberg(1991)]{Du91} Dubinski, J., \& Carlberg,
  R.~G.\ 1991, \apj, 378, 496

\bibitem[Evans et al.(2004)]{Ev04} Evans, N.~W.,
  Ferrer, F., \& Sarkar, S.\ 2004, PRD, 69, 123501

\bibitem[Evans et al.(2009)]{Ev09} Evans, N.~W., An, J., 
\& Walker, M.~G.\ 2009, \mnras, 393, L50 

\bibitem[Goerdt et al.(2006)]{Go06} Goerdt, T., Moore, B., 
Read, J.~I., Stadel, J., \& Zemp, M.\ 2006, \mnras, 368, 1073 

\bibitem[Irwin \& Hatzidimitriou(1995)]{Ir95} Irwin, M., \&
  Hatzidimitriou, D.\ 1995, \mnras, 277, 1354

\bibitem[Jardel 
\& Gebhardt(2012)]{Ja12} Jardel, J.~R., \& Gebhardt, K.\ 2012, \apj, 746, 89 

\bibitem[Kleyna et al.(2003)]{Kl03} Kleyna, J.~T., Wilkinson, M.~I.,
  Gilmore, G., \& Evans, N.~W.\ 2003, \apjl, 588, L21

\bibitem[{\L}okas(2009)]{Lo09} {\L}okas, E.~L.\ 2009, 
\mnras, 394, L102 

\bibitem[Macci{\`o} et al.(2007)]{Ma07} Macci{\`o}, A.~V., 
Dutton, A.~A., van den Bosch, F.~C., et al.\ 2007, \mnras, 378, 55 

\bibitem[Navarro et al.(1997)]{Na97} Navarro, J.~F., Frenk, 
C.~S., \& White, S.~D.~M.\ 1997, \apj, 490, 493 

\bibitem[Pontzen 
\& Governato(2012)]{Po12} Pontzen, A., \& Governato, F.\ 2012, \mnras, 421, 3464

\bibitem[Read \& Gilmore(2005)]{Re05} Read, J.~I., \&
  Gilmore, G.\ 2005, \mnras, 356, 107

\bibitem[S{\'a}nchez-Salcedo et al.(2006)]{SS06} 
S{\'a}nchez-Salcedo, F.~J., Reyes-Iturbide, J., 
\& Hernandez, X.\ 2006, \mnras, 370, 1829 

\bibitem[Strigari et al.(2010)]{St10} Strigari, L.~E., Frenk, C.~S.,
  \& White, S.~D.~M.\ 2010, \mnras, 408, 2364

\bibitem[Teyssier et al.(2012)]{Tey12} Teyssier, R., Pontzen, 
A., Dubois, Y., \& Read, J.\ 2012, arXiv:1206.4895 

\bibitem[Wilkinson et al.(2002)]{Wi02} Wilkinson, M.~I., 
Kleyna, J., Evans, N.~W., \& Gilmore, G.\ 2002, \mnras, 330, 778 

\bibitem[Walker et al.(2009a)]{Wa09a} Walker, M.~G.,
  Mateo, M., \& Olszewski, E.~W.\ 2009, \aj, 137, 3100

\bibitem[Walker et al.(2009b)]{Wa09} Walker, M.~G., Mateo, 
M., Olszewski, E.~W., et al.\ 2009, \apj, 704, 1274 

\bibitem[Walker et al.(2011)]{Wa11} Walker, M.~G., Combet, 
C., Hinton, J.~A., Maurin, D., \& Wilkinson, M.~I.\ 2011, \apjl, 733, L46 

\bibitem[Walker 
\& Pe{\~n}arrubia(2011)]{WaP11} Walker, M.~G., \& Pe{\~n}arrubia, J.\ 2011, \apj, 742, 20 

\bibitem[Wu(2007)]{Wu07} Wu, X.\ 2007, arXiv:astro-ph/0702233 

\end{thebibliography}
\end{document}